\documentclass[fleqn]{llncs}
\usepackage{latexsym}
\usepackage{amssymb,amsmath}
\usepackage{stmaryrd}
\usepackage{graphicx}
\usepackage{hyperref}
\usepackage{phonetic}
\usepackage{xargs}
\usepackage[pdftex,dvipsnames]{xcolor}
\usepackage{listings}

\newcommand{\be}{\begin{enumerate}}
\newcommand{\ee}{\end{enumerate}}
\newcommand{\bi}{\begin{itemize}}
\newcommand{\ei}{\end{itemize}}
\newcommand{\bc}{\begin{center}}
\newcommand{\ec}{\end{center}}
\newcommand{\bsp}{\begin{sloppypar}}
\newcommand{\esp}{\end{sloppypar}}

\newcommand{\sE}{\mbox{$\cal E$}}

\renewcommand{\phi}{\varphi}

\newcommand{\churchqe}{$\mbox{\sc ctt}_{\rm qe}$}
\newcommand{\churchuqe}{$\mbox{\sc ctt}_{\rm uqe}$}

\newcommand{\qzero}{${\cal Q}_0$}

\newcommand{\HOL}{$\mbox{\rm HOL}$}
\newcommand{\HL}{$\mbox{\rm HOL Light}$}
\newcommand{\HLQE}{$\mbox{\rm HOL Light QE}$}
\newcommand{\OCAML}{$\mbox{\rm OCaml}$}
\newcommand{\MMT}{$\mbox{\sc Mmt}$}

\newcommand{\mname}[1]{\mbox{\sf #1}}

\newcommand{\tarrow}{\rightarrow}

\usepackage[colorinlistoftodos,textsize=tiny]{todonotes}
\newcommandx{\unsure}[2][1=]{\todo[linecolor=red,backgroundcolor=red!25,bordercolor=red,#1]{#2}}
\newcommandx{\change}[2][1=]{\todo[linecolor=blue,backgroundcolor=blue!25,bordercolor=blue,#1]{#2}}
\newcommandx{\info}[2][1=]{\todo[linecolor=OliveGreen,backgroundcolor=OliveGreen!25,bordercolor=OliveGreen,#1]{#2}}
\newcommandx{\improvement}[2][1=]{\todo[linecolor=Plum,backgroundcolor=Plum!25,bordercolor=Plum,#1]{#2}}

\title{Biform Theories: Project Description\thanks{This research is supported by NSERC.}}

\author{Jacques Carette, William M. Farmer, and Yasmine Sharoda}

\institute{%
Computing and Software, McMaster University, Canada\\
\url{http://www.cas.mcmaster.ca/~carette}\\
\url{http://imps.mcmaster.ca/wmfarmer}\\[1.5ex]
12 June 2018
}

\begin{document}

\maketitle

\begin{abstract}
A \emph{biform theory} is a combination of an axiomatic theory and an
algorithmic theory that supports the integration of reasoning and
computation.  These are ideal for specifying and reasoning about
algorithms that manipulate mathematical expressions.  However,
formalizing biform theories is challenging as it requires the means to
express statements about the interplay of what these algorithms do and
what their actions mean mathematically.  This paper describes a
project to develop a methodology for expressing, manipulating,
managing, and generating mathematical knowledge as a network of biform
theories.  It is a subproject of MathScheme, a long-term project at
McMaster University to produce a framework for integrating formal
deduction and symbolic computation.
\end{abstract}

\noindent
We present the \emph{Biform Theories} project, a
subproject of MathScheme~\cite{CaretteEtAl11} (a long-term project
to produce a framework integrating formal
deduction and symbolic computation).

\section{Motivation}\label{sec:problem}

Type \texttt{2 * 3} into your favourite computer algebra system, press
enter, and you will receive (unsurprisingly) \texttt{6}. But what if
you want to go in the opposite direction?  Easy: you ask
$\texttt{ifactors(6)}$ in Maple or $\texttt{FactorInteger[6]}$ in
Mathematica.\footnote{Other computer algebra systems have similar
  commands.}  The Maple command \texttt{ifactors} returns a 2-element
list, with the first element the unit (\texttt{1} or \texttt{-1}) and
the second element a list of pairs (encoded as two-element lists) with
(distinct) primes in the first component and the prime's multiplicity
in the second. Mathematica's \texttt{FactorInteger} is similar, except
that it omits the unit (and thus does not document what happens for
negative integers).

This simple example illustrates the difference between a simple
computation \texttt{2 * 3} and a more complex \emph{symbolic} query,
factoring.  The reason for using lists-of-lists in both systems is
that multiplication and powering are both functions that evaluate
immediately in these systems.  So that factoring \texttt{6} cannot
just return \verb+2^1 * 3^1+, as that simply evaluates to \texttt{6}.
Thus it is inevitable that both systems must \emph{represent}
multiplication and powering in some other manner.  Because
\texttt{ifactors} and \texttt{FactorInteger} are so old, they are
unable to take advantage of newer developments in both systems, in
this case a feature to not immediately evaluate an expression but
leave it as a representation of a future computation.  Maple calls
this feature an \emph{inert form}, while in Mathematica it is a
\emph{hold form}.  Nevertheless, the need for representing future
computations was recognized early on: even in the earliest days of
Maple, one could do \texttt{5 \&\textasciicircum 256 mod 379} to
compute the answer without ever computing $5^{256}$ over the integers.
In summary, this example shows that in some cases we are interested in
\texttt{2 * 3} for its value and in other cases we are interested in
it for its syntactic structure.

A legitimate question would be: Is this an isolated occurrence, or a
more pervasive pattern? It is pervasive. It arises from the dichotomy
of being able to \emph{perform} computations and being able to
\emph{talk about} (usually to prove the correctness of) computations.
For example, we could represent (in Haskell) a tiny language of
arithmetic as
\begin{lstlisting}
data Arith = 
    Int Integer 
  | Plus Arith Arith 
  | Times Arith Arith
\end{lstlisting}
\noindent and an evaluator as
\begin{lstlisting}
eval :: Arith -> Integer
eval (Int x) = x
eval (Plus a b) = eval a + eval b
eval (Times a b) = eval a * eval b
\end{lstlisting}
\noindent whose ``correctness'' seems self-evident.  But what if we had
instead written
\begin{lstlisting}
data AA = TTT Integer | XXX AA AA | YYY AA AA

eval' :: AA -> Integer
eval' (TTT x) = x
eval' (XXX a b) = eval' a * eval' b
eval' (YYY a b) = eval' a + eval' b
\end{lstlisting}
how would we know if this implementation of \lstinline|eval'| is
correct or not?  The two languages are readily seen to be isomorphic.
In fact, there are clearly \emph{two} different isomorphisms. As the
symbols used are no longer mnemonic, we have no means to (informally!)
decide whether \lstinline|eval'| is correct.  Nevertheless,
\lstinline|Arith| and \lstinline|AA| both represent (trivial) embedded
\emph{domain specific languages} (DSLs), which are pervasively used in
computing. Being able to know that a function defined over a DSL is
correct is an important problem.

In general, both computer algebra systems (CASs) and theorem proving
systems (TPSs) manipulate \emph{syntactic representations} of
mathematical knowledge.  But they tackle the same problems in
different ways. In a CAS, it is a natural question to take a
polynomial $p$ (in some representation that the system recognizes as
being a polynomial) and ask to factor it into a product of
irreducible polynomials~\cite{von2003modern}.  The algorithms to do
this have gotten extremely sophisticated over the
years~\cite{vanhoeij2002}.  In a TPS, it is more natural to prove that
such a polynomial $p$ is equal to a particular factorization, and
perhaps also prove that each such factor is irreducible. Verifying
that a given factorization is correct is, of course, easy. Proving
that factors are irreducible can be quite hard. And even though
CASs obtain
information that would be helpful to a TPS towards such a proof, that
information is usually not part of the output. Thus while some
algorithms for factoring do produce irreducibility
\emph{certificates}, which makes proofs straightforward, these are
usually not available. And the complexity of the algorithms (from an
engineering point of view) is sufficiently daunting that, as far as we
know, no TPS has re-implemented them.

Given that both CASs and TPSs ``do mathematics'', why are they so 
different? Basically because a CAS is based around
\emph{algorithmic theories}, which are collections of symbolic computation
algorithms whose correctness has been established using pen-and-paper
mathematics, while a TPS is based around \emph{axiomatic theories},
comprised of signatures and axioms, but nevertheless representing the
``same'' mathematics. In a TPS, one typically proves theorems, formally.
There is some cross-over: some TPSs (notably Agda and Idris) are closer
to programming languages, and thus offer the real possibility of mixing
computation and deduction. Nevertheless, the problem still exists: how
does one verify that a particular function implemented over a representation
language carries out the desired computation?

What is needed is a means to \emph{link} together axiomatic theories
and algorithmic theories such that one can state that some ``symbol
manipulation'' corresponds to a (semantic) function defined axiomatically?
In other words, we want to know that a \emph{symbolic computation}
performed on representations performs the same computation as an abstract
function defined on the \emph{denotation} of those representations.
For example, if we ask to integrate a particular expression $e$, we would like
to know that the system's response will in fact be an expression representing
an integral of $e$ --- even if the formal definition of integration uses an
infinitary process.

These kinds of problems are pervasive: not just closed-form symbolic
manipulations, but also SAT solving, SMT solving, model checking,
type-checking of programs, and most manipulations of DSL terms, are
all of this sort.  They all involve a mixture of computation and
deduction that entwine syntactic representations with semantic
conditions.

In the next section we will introduce the notion of a \emph{biform
  theory} that is a combination of an axiomatic theory and an
algorithmic theory so that we can define and reason about symbolic
computation in the same setting.

\section{Background Ideas}

A \emph{transformer} is an algorithm that implements a function $\sE^n
\tarrow \sE$ where $\sE$ is a set of expressions.  The expressions
serve as data that can be manipulated.  Different kinds of expressions
correspond to different data representations.  Transformers can
manipulate expressions in various ways.  Simple transformers, for
example, build bigger expressions from pieces, select components of
expressions, or check whether a given expression satisfies some
syntactic property.  More sophisticated transformers manipulate
expressions in mathematically meaningful ways.  We call these kinds of
transformers \emph{syntax-based mathematical algorithms
  (SBMAs)}~\cite{Farmer13}.  Examples include algorithms that apply
arithmetic operations to numerals, factor polynomials, transpose
matrices, and symbolically differentiate expressions with variables.
The \emph{computational behavior} of a transformer is the relationship
between its input and output expressions.  When the transformer is an
SBMA, its \emph{mathematical meaning}%
\footnote{Computer scientists would call this \emph{denotational semantics}
rather than \emph{mathematical meaning}.} is the relationship between the
mathematical meanings of its input and output expressions.

A \emph{biform theory} $T$ is a triple $(L,\Pi,\Gamma)$ where $L$ is a
language of some underlying logic, $\Pi$ is a set of transformers that
implement functions on expressions of $L$, and $\Gamma$ is a set of
formulas of
$L$~\cite{CaretteFarmer08,Farmer07b,FarmerMohrenschildt03}.  $L$
includes, for each transformer $\pi \in \Pi$, a name for the function
implemented by $\pi$ that serves as a name for $\pi$.  The members of
$\Gamma$ are the \emph{axioms} of $T$.  They specify the meaning of
the nonlogical symbols in $L$ including the names of the transformers
of $T$.  In particular, $\Gamma$ may contain specifications of the
computational behavior of the transformers in $\Pi$ and of the
mathematical meaning of the SBMAs in $\Pi$.  A formula in $\Gamma$
that refers to the name of a transformer $\pi \in \Pi$ is called a
\emph{meaning formula} for $\pi$.  The transformers in $\Pi$ may be
written in the underlying logic or in an programming language external
to the underlying logic.  We say $T$ is an \emph{axiomatic theory} if
$\Pi$ is empty and an \emph{algorithmic theory} if $\Gamma$ is empty.

\begin{example}
Let $R_{\rm ax} = (L,\Gamma)$ be a first-order axiomatic theory of a
ring with identity.  The language $L$ contains the usual constants (0
and 1), function symbols ($+$ and $*$), and predicate symbols ($=$),
and $\Gamma$ contains the usual axioms.  The terms of $L$, which are
built from 0, 1, and variables by applying $+$ and $*$, have the form
of multivariate polynomials.  Thus algorithms that manipulate
polynomials --- that normalize a polynomial, factor a polynomial, find
the greatest common divisor of two polynomials, etc. --- would
naturally be useful for reasoning about the expressions of $R_{\rm
  ax}$.  Let $\Pi$ be a set of such transformers on the terms in $L$,
$L'$ be an extension of $L$ that includes vocabulary for naming and
specifying the transformers in $\Pi$, and $\Gamma'$ contain meaning
formulas for the transformers in $\Pi$ expressed in $L'$.  Then
$R_{\rm bt} = (L',\Pi,\Gamma \cup \Gamma')$ is a biform theory for
rings with identity.  It would be very challenging to express $R_{\rm
  bt}$ in ordinary first-order logic; the meaning formulas in
$\Gamma'$ would be especially difficult to express.  Notice that
$R_{\rm alg} = (L',\Pi)$ is algorithmic theory of multivariate
polynomials with constants 0 and 1.
\end{example}

Formalizing a biform theory in the underlying logic requires
infrastructure for reasoning about the expressions manipulated by the
transformers as syntactic entities.  This infrastructure provides a
basis for \emph{metareasoning with reflection}~\cite{Farmer18}.
There are two main approaches to build such an
infrastructure~\cite{Farmer13}.  The \emph{local approach} is to
produce a deep embedding of a sublanguage $L'$ of $L$ that includes
all the expressions manipulated by the transformers of $\Pi$.  The
\emph{global approach} is to replace the underlying logic of $L$ with
a logic such as \churchqe~\cite{Farmer18} that has an inductive
type of \emph{syntactic values} that represent the expressions in $L$
and global quotation and evaluation operators.  A third approach,
based on ``final tagless'' embeddings~\cite{CaretteKS09}, has not yet
been attempted as most logics do not have the necessary infrastructure
to abstract over type constructors.

A complex body of mathematical knowledge can be represented in
accordance with the \emph{little theories method}~\cite{FarmerEtAl92b}
(or even the \emph{tiny theories method}~\cite{CaretteOConnorTPC})
as a \emph{theory graph}~\cite{Kohlhase14} consisting of axiomatic
theories as nodes and theory morphisms as directed edges.  A
\emph{theory morphism} is a meaning-preserving mapping from the
formulas of one axiomatic theory to the formulas of another.  The
theories --- which may have different underlying logics --- serve as
abstract mathematical models, and the morphisms serve as information
conduits that enable theory components such as definitions and
theorems to be transported from one theory to
another~\cite{BarwiseSeligman97}.  A theory graph enables mathematical
knowledge to be formalized in the most convenient underlying logic at
the most convenient level of abstraction using the most convenient
vocabulary.  The connections made by the theory morphisms in a theory
graph then provide the means to find this knowledge and apply it in
other contexts.

A \emph{biform theory graph} is a theory graph whose nodes are biform
theories. Having the same benefits as theory graphs of axiomatic
theories, biform theory graphs are well suited for representing
mathematical knowledge that is expressed both axiomatically and
algorithmically.
  
Our previous work on mechanized mathematics systems and on related
technologies has taught us that such a graph of biform theories 
really should be a central component of any future systems for
mathematics. We will expand on the objectives of the project and
its current state. At the same time, additional pieces of the project
beyond what is motivated above (but is motivated by previous and
related work) will be weaved in as appropriate.

\section{Project Objectives}

The primary objective of the Biform Theories project is:

\bi

  \item[] \textbf{Primary.} Develop a methodology for expressing,
    manipulating, managing and generating mathematical knowledge 
    as a biform theory graph.

\ei

\noindent
Our strategy for achieving this is to break down the problem
into the following subprojects:

\bi

  \item[]\textbf{Logic} Design a logic \mname{Log} which is a version
    of simple type theory~\cite{Farmer08} with an inductive type of
    syntactic values, a global quotation operator, and a global
    evaluation operator.  In addition to a syntax and semantics,
    define a proof system for \mname{Log} and a notion of a theory
    morphism from one axiomatic theory of \mname{Log} to another.
    Demonstrate that SBMAs can be defined in \mname{Log} and that
    their mathematical meanings can be stated, proved, and
    instantiated using \mname{Log}'s proof system.

\medskip

  \item[]\textbf{Implementation} Produce an implementation \mname{Impl} of
    \mname{Log}.  Demonstrate that SBMAs can be defined in
    \mname{Impl} and that their mathematical meanings can be stated
    and proved in \mname{Impl}.

\medskip

  \item[]\textbf{Transformers} Enable biform theories to be defined in
    \mname{Impl}.  Introduce a mechanism for applying transformers
    defined outside of \mname{Impl} to expressions of \mname{Log}.
    Ensure that we know how to write meaning formulas for such
    transformers. Some transformers can be automatically generated
    --- investigate the scope of this, and implement those which
    are feasible.

\medskip

  \item[]\textbf{Theory Graphs} Enable theory graphs of biform
    theories to be defined in \mname{Impl}.  Use combinators to ease
    the construction of large, structured biform theory graphs.
    Introduce mechanisms for transporting definitions, theorems, and
    transformers from a biform theory $T$ to an instance $T'$ of $T$
    via a theory morphism from $T$ to $T'$. Some theories (such as
    theories of homomorphisms and term languages) can be and thus
    should be automatically generated.

\medskip

  \item[]\textbf{Generic Transformers} Design and implement in \mname{Impl} a
    scheme for defining generic transformers in a theory graph $T$
    that can be specialized, when transported to an instance $T'$ of
    $T$, using the properties exhibited in $T'$.

\ei

\section{Work Plan Status}

The work plan is to pursue the five subprojects described above more
or less in the order of their presentation.  Here we describe the
parts of the work plan that have been completed as well as the parts
that remain to be done.

\subsection*{Logic with Quotation and Evaluation}

This subproject is largely complete.  We have developed
{\churchqe}~\cite{Farmer18}, a version of Church's type
theory~\cite{Church40} with global quotation and evaluation operators.
(Church's type theory is a popular form of simple type theory with
lambda notation.)  The syntax of {\churchqe} has the machinery of
{\qzero}~\cite{Andrews02}, Andrews' version of Church's type theory
plus an inductive type $\epsilon$ of syntactic values, a partial
quotation operator, and a typed evaluation operator.  The semantics of
{\churchqe} is based on Henkin-style general models~\cite{Henkin50}.
The proof system for {\churchqe} is an extension of the proof system
for {\qzero}.

We show in~\cite{Farmer18} that {\churchqe} is suitable for defining
SBMAs and stating, proving, and instantiating their mathematical
meanings.  In particular, we prove within the proof system for
{\churchqe} the mathematical meaning of a symbolic differentiation
algorithm for polynomials.

We have also defined {\churchuqe}~\cite{Farmer17}, a variant of
{\churchqe} in which undefinedness is incorporated in {\churchqe}
according to the traditional approach to
undefinedness~\cite{Farmer04}.  Better suited than {\churchqe} as a
logic for interconnecting axiomatic theories, we have defined in
{\churchuqe} a notion of a theory morphism~\cite{Farmer17}.

\subsection*{Implementation of the Logic}

We have produced an implementation of {\churchqe} called
{\HLQE}~\cite{CaretteFarmerLaskowski18} by modifying
{\HL}~\cite{Harrison09}, an implementation of the {\HOL} proof
assistant~\cite{GordonMelham93}.  {\HLQE} provides a built-in global
infrastructure for metareasoning with reflection.  Over the next
couple years we plan to test this infrastructure by formalizing a
variety of SBMAs in {\HLQE}.

Building on the experience we gain in the development of {\HLQE}, we
would like to create an implementation of {\churchqe} in
{\MMT}~\cite{RabeKohlhase13} that is well suited for general use and
has strong support for building theory graphs.  We will transfer to
this {\MMT} implementation the most successful of the ideas and
mechanisms we develop on the three subprojects that follow using
{\HLQE}.

\subsection*{Biform Theories, Transformers, and Generation}

Implementation of biform theories in {\HLQE} has not yet started, but
we expect that it will be straightforward, as will the application of
external transformers.  External transformers implemented in {\OCAML}
(or in languages reachable via {\OCAML}'s foreign function interface)
can be linked in as well.

The most difficult part of this subproject will be adequate renderings
of \emph{meaning formulas} that express the mathematical meaning of
transformers. We do have some
experience~\cite{CaretteFarmer17,CaretteFarmerSorge07} creating biform
theories.  The exploration and implementation of automatic generation
of transformers has started.

\subsection*{Biform Theory Graphs}

In~\cite{CaretteFarmer17}, we developed a case study of a biform
theory graph consisting of eight biform theories encoding natural number
arithmetic.  We produced partial formalizations of this test
case~\cite{CaretteFarmer17} in {\churchuqe}~\cite{Farmer17} using the
global approach for metareasoning with reflection, and in
Agda~\cite{Norell07,Norell09} using the local approach.  After we have
finished with the previous two subprojects, we intend to formalize this
in {\HLQE} as well.

In~\cite{CaretteOConnorTPC}, we developed combinators for combining
theory presentations. There is no significant difference between
axiomatic and biform theories with respect to the semantics of
these combinators, and we expect that these will continue to work
as well as they did in~\cite{MathSchemeExper}. There, we also
experimented with some small-scale theory generation, which worked
well. This subproject will also encompass the implementation of
\emph{realms}~\cite{CaretteEtAl14}.  We also hope to make some
inroads on \emph{high level theories}~\cite{CaretteFarmer08}.

\subsection*{Generic, Specializable Transformers}

Through substantial previous work~%
\cite{CaretteKS09,Carette06,CaElSm11,MathSchemeExper,CaretteKiselyov2005,CaretteKiselyov11,CaretteKucera07,CaretteKucera11,carette2016simplifying,KuceraCarette06,Larjani13,narayanan2016probabilistic}
on code generation and code manipulation, it has become quite clear that
quite a lot of mathematical code can be automatically generated.
One of the most successful techniques is \emph{instantiation},
whereby a single, generic algorithm exposes a series of
\emph{design choices} that must be explicitly instantiated to produce
specialized code. By clever choices of design parameters, and through the
use of partial evaluation, one can thus produce highly optimized code
without having to hand-write such code.

\section{Related Work}

Directly related is~\cite{KohlhaseManceRabe13} whose authors also work
with biform theory graphs. Michael Kohlhase and Florian Rabe and their
students are actively working on related topics. As a natural
progression, we (the authors of this paper) have started actively
collaborating with them, under the name of the \emph{Tetrapod
  Project}.

One of the crucial features for supporting the interplay between
syntax and semantics is \emph{reflection}, which has a long history
and a deep literature. The interested reader should read the 
thorough related work section in~\cite{Farmer18} for more details.

There are substantial developments happening in some systems, most
notably Agda~\cite{Norell07,Norell09}, Idris~\cite{Brady13} and
Lean~\cite{Lean} that we are paying particularly close attention to.  This includes
quite a lot of work on making reflection practical~%
\cite{Christiansen:2016,Christiansen:2014,ebner2017metaprogramming,VanDerWalt12}.

On the more theoretical side, \emph{homotopy type theory}~\cite{hottbook}
is rather promising.  However quite a bit of research still needs to be
done to make these results practical. Of particular note is the issue
that theories that deal directly with syntax seem to clash with the
notion of a \emph{univalent universe}, which is central to homotopy
type theory.

\section{Conclusion}

Building mechanized mathematics systems is a rather complex engineering
task. It involves creating new science --- principally through the creation
of logics which can support reasoning about syntax.  It also involves significant
new engineering --- both on the systems side, where \emph{knowledge management}
is crucial to reduce the \emph{information duplication} inherent in a
naive implementation of mathematics, and on the usability front, where
users do not, and should not, care about all the infrastructure that 
developers need to create their system. Current systems tend to expose
this infrastructure, thus creating an additional burden for casual users
who may well have a simple task to perform.

The \emph{Biform Theories} project is indeed about infrastructure that
we believe is essential to building large-scale mechanized mathematics
systems. And yes, we do believe that eventual success would imply that
casual users of such a system never hear of ``biform theories''.

\section*{Acknowledgments} 

This research was supported by NSERC.  The authors would like to thank
the referees for their comments and suggestions.

\bibliography{imps}
\bibliographystyle{splncs04}


\end{document}